\documentclass[12pt]{aa}
\usepackage[english]{babel}
\usepackage{graphicx}
\usepackage{longtable}

\onecolumn
\begin{document}

\title{High-resolution optical spectroscopy of the post-AGB supergiant V340 Ser \\  (=IRAS\,17279$-$1119)}
\author{V.G. Klochkova, V.E. Panchuk, N.S. Tavolzhanskaya,  and M.V. Yushkin \\
{\small \email{valenta@sao.ru}}}

\institute{Special Astrophysical Observatory RAS, Nizhnij Arkhyz,  369167 Russia \\}
\date{\today}


\abstract{Some evidences of wind variability and velocity stratification in the extended atmosphere has been
found in the spectra of the supergiant V340\,Ser (=IRAS\,17279$-$1119) taken at the 6-m BTA telescope
with a spectral resolution R$\ge$60000. The H$\alpha$ line has a P\,Cyg profile whose absorption component 
(V=+34\,km/s) is formed in the upper layers of the expanding atmosphere close to the circumstellar environment.
For four dates the mean velocity has been derived from the positions of 300--550 symmetric metal
absorptions with an accuracy better than $\pm0.1$\,km/s: V$_{\sun}$\,=\,59.30, 60.09, 58.46, and 55.78\,km/s.
A lot of low-excitation metal lines have an inverse P\,Cyg profile. The mean positions of their emission
components, V$_{\sun}$=\,46.3$\pm$0.4\,km/s, differ systematically from the velocity inferred from symmetric
absorptions, suggesting the presence of a velocity gradient in the supergiant extended atmosphere. The
multicomponent profile of the NaI\,D lines contains the interstellar, V$_{\sun}= -11.2$\,km/s, and circumstellar,
V$_{\sun}$=+10\,km/s, components and the component forming in the upper atmospheric layers, 
V$_{\sun}$=+34.0\,km/s. The mean velocity from 20--30 diffuse interstellar bands (DIBs) identified in the spectra,
V$_{\sun}$(DIBs)\,=$-11.6\pm0.2$\,km/s, agrees with the velocity from interstellar NaI and KI components. 
The equivalent width of the oxygen triplet W(OI\,7774)\,=\,1.25\,\AA{} corresponds to an absolute magnitude of the
star Mv$\approx -4.6^m$, which, taking into account the total (interstellar\,+\,circumstellar) extinction, 
leads to a distance to the star d$\approx$2.3\,kpc.
}
\titlerunning{\small Optical spectroscopy of V340\,Ser}
\authorrunning{\small Klochkova et al.}

\maketitle

\section{INTRODUCTION}

V340\,Ser, the central star of the infrared (IR) source IRAS 17279$-$1119, was entered into a list
of stars with IR excesses by Oudmaijer et al.~(1992). Later, having studied the Hubble Space Telescope
data for a large sample of IR sources, Si\`odmiak et al.~(2008) showed IRAS\,17279$-$1119 to have
a point image and classified V340 Ser as a postasymptotic giant branch (hereafter post-AGB) star.
These authors pointed out the presence of a near-IR excess, suggesting the presence of warm 
circumstellar dust. Highly evolved stars with initial masses in the range $2\div8 \mathcal{M}_{\sun}$ 
are  observed at the post-AGB stage.

According to Bl\"ocker (1995a), at the preceding AGB evolutionary stage these stars are observed as 
red supergiants with effective temperatures Teff$\approx 3000\div4500$\,K. The AGB stage for stars 
with the above masses is the final evolution phase with nucleosynthesis in the stellar cores. 
The interest in AGB stars and their closest descendants stems primarily from the fact that in 
the interiors of these stars, which are at a short evolutionary stage, there are physical
conditions for the synthesis of heavy-metal nuclei and the dredge-up of nuclear reaction 
products into the stellar atmosphere and then into the circumstellar and interstellar medium. 
Due to these processes, AGB stars with initial masses below $3\div4\mathcal{M}_{\sun}$  are the
main suppliers  of all the elements heavier than iron synthesized through the s-process, 
which basically consists in slow (compared to $\beta$ decay) neutronization of nuclei. Fe nuclei serve 
as initial nuclei for the chain of s-process reactions. In stars with initial mass below 
$3\div4\mathcal{M}_{\sun}$  the required neutron ﬂux is provided by the $^{13}C(\alpha,n)^{16}O$ 
reaction, while in the case of more massive stars, with initial mass above $4\div5\mathcal{M}_{\sun}$, 
the analogous  reaction proceeds on  $\rm ^{22}Ne$ nuclei. It should be emphasized that these more
massive AGB stars can also be the sources of lithium. The details of the evolution of stars near the AGB
and the results of present-day calculations of the synthesis and dredge-up of elements are presented
in Herwig~(2005), di Criscienzo et al.~(2016), and Liu et al.~(2018).

The spectral type F2--3\,II is given for the supergiant V340\,Ser in the SIMBAD database. This
supergiant lies outside the Galactic plane, which already suggests that it probably belongs to 
low mass evolved stars. By now the star has been studied by both photometric and spectroscopic methods. 
A number of authors (in particular, Kazarovets et al.~2000; de~Smedt et al.~2016) classify 
V340\,Ser as an RV\,Tau variable. The post-AGB stars of this category are located on the 
Hertzsprung--Russell diagram in (or near) the instability strip and have near-IR excesses, 
which serves as evidence for the presence of warm circumstellar dust that has not yet
detached from the photospheric layers of the star.

The abundances of chemical elements in the atmosphere of V340\,Ser have been studied repeatedly
and quite reliably. Based on high-resolution spectra, but with a moderate S/N ratio, 
Arellano Ferro et al.~(2001) derived the model atmosphere parameters for this star: 
Teff\,=\,7300\,K, surface gravity log\,g\,=\,1.25, and microturbulence $\xi_t$\,=\,4.6\,km/s.
They found a lowered metallicity, [Fe/H]$_{\sun}$=$-0.60$, an overabundance of heavy s-process elements, 
and a carbon-to-oxygen abundance ratio C/O$\approx$1. As a result, these authors supported the 
previous conclusion (van Winckel, 1997) that the star belongs to the post-AGB stars that have 
passed  the third dredge-up. 

The paper by Arellano Ferro et al.~(2001) is also important in that its authors pointed out the
absence of selective depletion of chemical elements in the atmosphere of this star, which usually 
distorts the elemental abundances in the atmospheres of related post-AGB stars (for examples, 
see Klochkova~(1995), Giridhar et al.~(2000), Rao et al.~(2012), Klochkova and Tavolzhanskaya~(2019)).

Based on high-resolution spectra, Rao et al. (2012) determined the model parameters and detailed 
atmospheric chemical composition for a sample of post-AGB stars, including V340\,Ser. The model
parameters adopted by them for this star are, i.e., Teff\,=\,7300$\pm$150\,K, surface gravity 
log\,g\,=\,2.25$\pm$0.25, and microturbulence $\xi_t$=\,4.7$\pm$0.25\,km\,s, agree, within 
the error limits, with the previously derived parameters from Arellano Ferro et al.~(2001).
At a lowered  metallicity, [Fe/H]$_{\sun}$=$-0.60$, a nontrivial atmospheric chemical composition 
was found for V340\,Ser: moderate, but significant enhancements of $\alpha$-process elements, 
$\rm [\alpha/Fe]$=+0.44, and heavy s-process metals, $\rm [s/Fe]$=+0.69. 

Later, using VLT+UVES spectra,  de Smedt et al.~(2016) again confirmed the parameters and chemical 
abundances of V340\,Ser. The main goal of their study, the search for evidence of lead in 
the atmospheres of the program stars, was not achieved, because for 14 post-AGB stars with 
heavy-metal-enriched atmospheres the authors obtained only upper limits for the abundance
of this element. The paper of these authors is important in our study of the optical spectra for
V340\,Ser, because, having performed a spectroscopic monitoring with the HERMES spectrograph of the
1.2-m telescope, de Smedt et al.~(2016) concluded that this star is a binary by determining its 
orbital parameters. The period found by them is P\,=\,365~days, the systemic velocity is 
$\gamma=56.8\pm0.5$ km/s, and the velocity amplitude is $\Delta Vr=7.92\pm0.44$\,km/s.

The behavior of the photometric parameters for V340\,Ser has also been studied reliably. Having
performed a 7-year-long UBV monitoring of the star, Arkhipova et al.~(2011) concluded that its brightness
is variable with an amplitude $\Delta V\approx 0.2^m$ typical of post-AGB stars and a fundamental 
period P\,= 89.6~days. Appending the ASAS data distributed more uniformly in time to their analysis, 
Arkhipova et al.~(2011) determined the amplitude, $\Delta V\approx 0.3^m$, and the fundamental 
period, P\,= 89.8~days. 

In this paper we present the results of our analysis of the optical spectra for V340\,Ser taken at 
the 6\,meter BTA telescope in 2018--2020.  The main goal of our work is the search for spectral 
peculiarities  and their variability with time. In Section~2 we briefly describe the methods of 
observations and data analysis. In Section~3 we present the results by comparing them with the 
previously published ones and in Section~3 we provide our conclusions.

\section{OBSERVATIONS, DATA REDUCTION, AND ANALYSIS OF SPECTRA}

\begin{table*}[ht]
\medskip
\caption{Results of our measurements of the heliocentric radial velocity V$_{\sun}$ in the spectra of V340\,Ser 
based on sets of lines of various types. The number of features of various types used is given in parentheses. 
The positions of the absorption components forming outside the stellar atmosphere are given for the NaI~D lines
V$_{\sun}$, km/s }
\begin{tabular}{ c|  c|  c|  c|  c|  l|  r| c }
\hline
JD & \multicolumn{7}{c}{\small  $V_{\sun}$, km/s} \\  
\cline{2-8}
$-$2458000 & Metal  & Metal& H$\alpha$(abs)& H$\beta$&\hspace{0.3cm}NaI &KI & DIBs \\  
           &absorptions&emissions  & H$\alpha$(em) &         &    &   &   \\   
\hline   
213.5 &59.30           &45.0           &34.2   &53.6  &$-11.3$,\,10.0,\,33.7 &$-9.2$ &$-11.9$ \\ 
      &$\pm$0.05\,(557)& $\pm$0.2\,(59)&101.7  &      &                      &       &$\pm$0.2\,(23)    \\    
\hline
220.4 &60.09           &46.0           &34.9   & 54.8 &$-11.2$,\,10.2,\,33.4 &$-10.0$& $-11.3$ \\  
      &$\pm$0.05\,(459)& $\pm$0.3\,(43)&100.1  &      &                      &       & $\pm$0.2\,(32)   \\   
\hline
574.4 &58.46           &48.0           &34.5   &53.9  &$-11.4$,\,10.3,\,35.0&$-10.7$ &$-12.1$ \\ 
      &$\pm$0.06\,(283)&$\pm$0.4 (49)  &101.2  &      &                      &       & $\pm0.3$\,(18)  \\ 
\hline
924.2 &55.78           &46.0           &33.6   &49.3  &$-11.1$,\hspace{0.3cm}9.1,\,32.8&$-9.3$ &$-11.3$ \\ 
      &$\pm$0.06\,(408)&$\pm$0.3 (28)  &97.5   &      &                      &       & $\pm0.3$\,(15)  \\ 
\hline
\end{tabular}   
\label{velocity}
\end{table*}    

The spectra of V340\,Ser were taken with the  Nasmyth echelle spectrograph (NES) (Panchuk
et al.~2017) at the 6-m BTA telescope. The dates of observations of the star are listed in Table\,\ref{velocity}. 
At these dates the NES was equipped with a large format 4608$\times$2048 CCD  with a pixel size of
0.0135$\times$0.0135\,mm, the readout noise is 1.8\,e$^{-}$.  The recorded spectral range is 
$\Delta\lambda$\,=\,470--778\,nm. To reduce the light losses without any loss of the spectral
resolution, the NES is equipped with an image slicer forming three slices. Each spectral order 
in the 2-dimensional image of the spectrum is repeated three times with a shift along the  cross  
dispersion. The spectral resolution is $ge 60000$, 
the signalto-noise ratio is S/N$>$100, it changes along the echelle order from 100 to 150.

The one-dimensional data were extracted from the 2-dimensional echelle spectra with the modified
(taking into account the peculiarities of the NES echelle frames) ECHELLE context of the MIDAS
software package. The details of the procedure were described by Yushkin and Klochkova~(2005). The
cosmic-ray particle hits were removed by a median averaging of two spectra taken successively 
one by one. The wavelength calibration was carried out using the spectra of a hollow-cathode Th-Ar lamp. 
The entire subsequent reduction, including the photometric and positional measurements, was performed 
with the up-to-date version of the DECH20t code developed by Galazutdinov~(1992). Note that this code
traditionally used by us to reduce the spectra allows the radial velocities for individual features in the line
proﬁles to be measured. The systematic measurement errors of the heliocentric velocities V$_{\sun}$  
estimated from sharp interstellar Na\,I components do not exceed 0.25\,km/s (from one line); 
the random errors for shallow absorptions are $\approx 0.5$\,km/s, the mean value per line. 
Thus, for the average values of V$_{\sun}$ in Table\,\ref{velocity} the random errors are 0.05$\div$0.3\,km/s, 
depending on the number of measured lines. We identified the features in the spectrum of V340\,Ser using the atlas
of the optical spectrum for the canonical post-AGB star HD\,56126 (=IRAS\,07134+1005, Sp\,=\,F5\,Iab)
published by Klochkova et al.~(2007a), which is based on the observational data of the 6-m BTA 
telescope in combination with the NES.

\section{RESULTS}

\subsection{Peculiarities of the optical spectrum for V340 Ser and radial velocity stratification}

\begin{figure}[bpt!!!] 
\includegraphics[angle=0,width=0.5\textwidth, bb=10 70 560 680,clip]{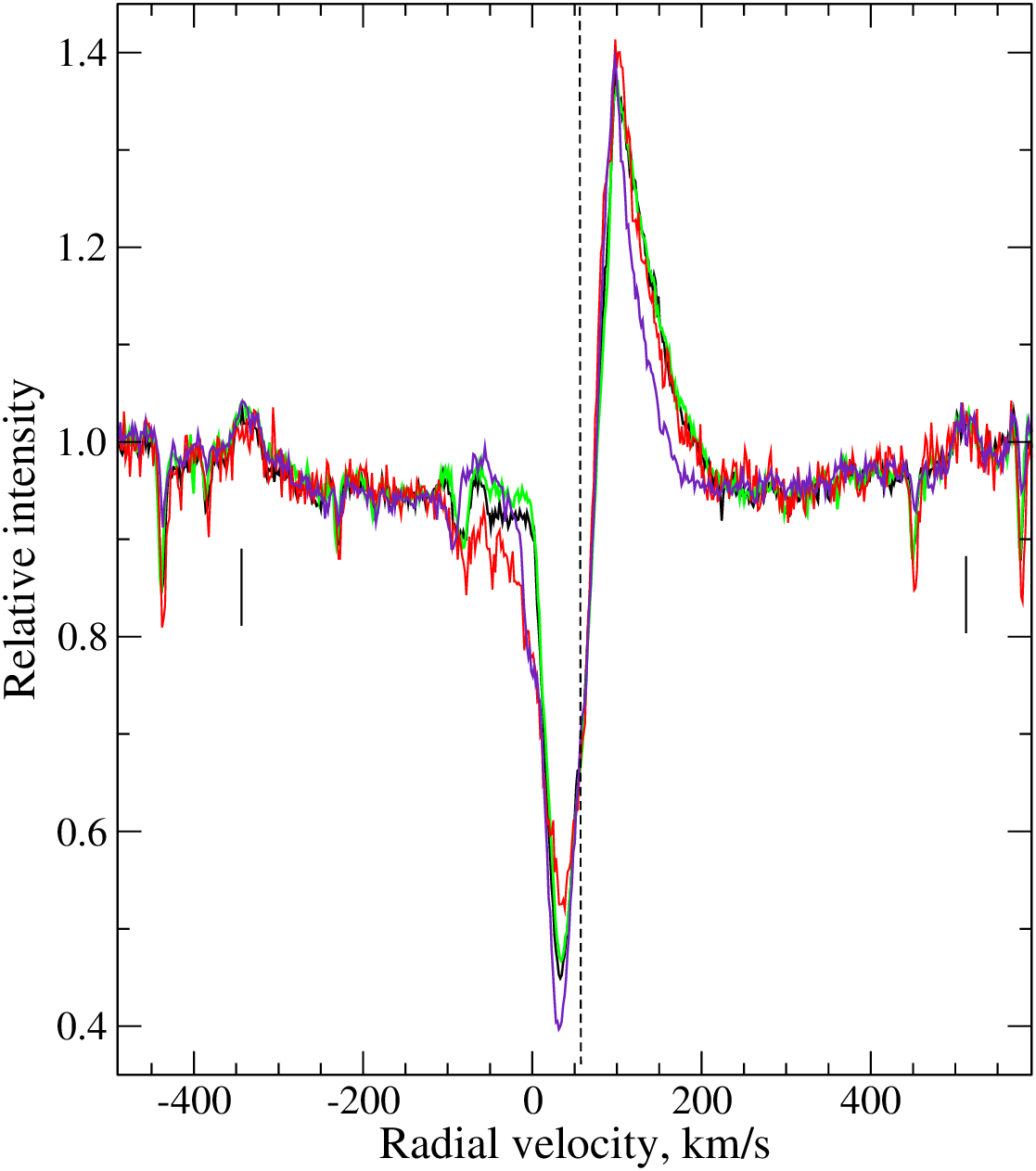}
\caption{H$\alpha$ profile in relative intensity -- radial velocity coordinates in the spectra of  V340\,Ser 
taken at different dates: April 5, 2018 (black line), April 11, 2018 (green line), March 31, 2019 (red line), 
and April 2, 2020 (indigo line). The two solid vertical bars indicate the positions of the TiI\,6554.23\,\AA{} 
and CaI\,6572.80\,\AA{} emissions. The vertical line marks the systemic velocity
$\gamma=56.8\pm0.5$\,km/s from de Smedt et al.~(2016). }
\label{Halpha-var}
\end{figure}

\begin{figure}[bpt!!!] 
\includegraphics[angle=0,width=0.5\textwidth, bb=10 70 560 680,clip]{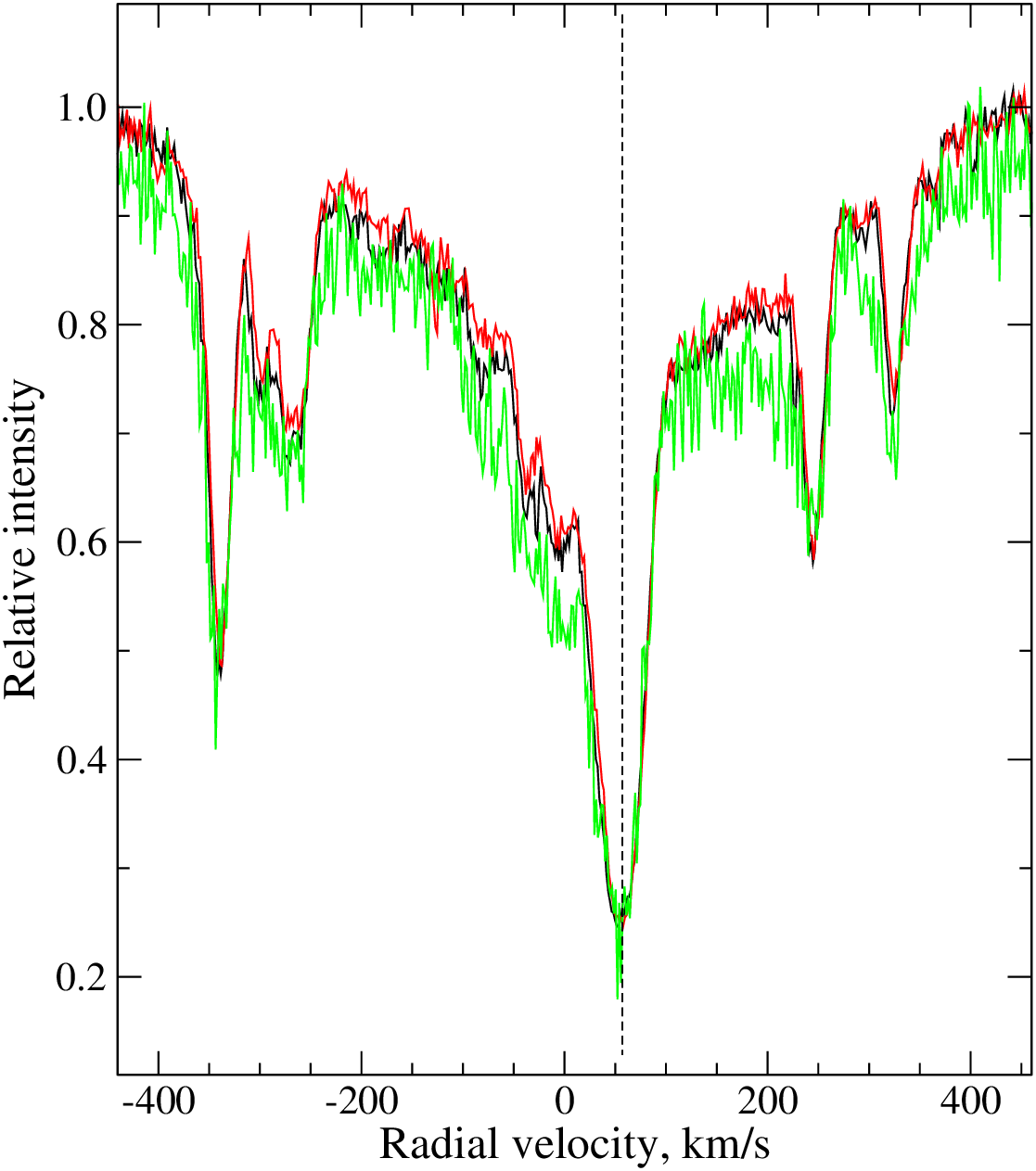}
\caption{H$\beta$ profile in the spectra of V340\,Ser taken at different dates: April 5, 2018 (black line), 
April 11, 2018 (green line), and March 31, 2019 (red line). The vertical line marks the systemic velocity
$\gamma=56.8\pm0.5$\,km/s from de Smedt et al.~(2016). }
\label{Hbeta}
\end{figure}

The spectral anomalies inherent in RV Tau stars are the emission components 
of the complex and variable (with phase) H$\alpha$ profiles, the HeI~5876\,\AA{} 
emission, the splitting of selected absorptions, the multicomponent profiles of 
the NaI~D lines, and the weak metal emissions appearing at certain times of the cycle. 
The spectra of the prototype stars RV\,Tau, U\,Mon, and AC\,Her studied by Baird~(1982, 1984), 
Pollard et al.~(1997), Klochkova and Panchuk (1998), and Kipper and Klochkova~(2013) serve 
as examples. Both the absorption splitting and the emission in the HeI\,5876\,\AA{} line 
are absent in our spectra of V340 Ser.  The optical spectrum of this star, which
largely corresponds to the expected spectrum of an F2--3 supergiant, contains 
some of the peculiarities listed above. First, the H$\alpha$ profile is complex, which
is typical of post-AGB stars. The profile of this line presented in Fig.\,\ref{Halpha-var} in 
relative intensity -- radial velocity coordinates includes the broad absorption wings on
which the absorption and emission components are superimposed. As follows from 
Fig.\,\ref{Halpha-var}, the positions of both components do not change with time, but a weak variability 
of their intensity can be noted. The change in the shape of the blue wing of the absorption
component is more pronounced, reflecting a change in the outflow velocity. A large difference 
between the velocities corresponding to the cores of the emission and absorption components 
in the H$\alpha$ profile should be emphasized. 
As can be seen from the data in  Table~\ref{velocity} and Fig.\,\ref{Halpha-var}, this difference is 
about 65\,km/s and exceeds the same parameter for the RV\,Tau stars studied by Pollard et al.~(1997), 
including the star AD\,Aql, in the spectrum of which the H$\alpha$ profile is
close to the one recorded by us in the spectrum of V340\,Ser, by many times. As can be seen 
from Fig.\,\ref{Hbeta}, the H$\beta$ profile is a purely absorption one, it
contains no visible emission features. As follows from Table\,\ref{velocity}, the position 
of the H$\beta$ core does not coincide quite closely with the positions of the atmospheric
metal absorptions, it is blueshifted by 5--6\,km/s, suggesting a stratification of the 
velocity pattern.

The circumstellar envelopes of evolved stars are often the sources of molecular and maser emission,
which allows their systemic velocity to be reliably recorded. However, for V340\,Ser there is 
no information about any envelope features in the radio band.
The optical spectrum of this star does not contain any evidence of molecules either and, 
therefore, we use  $\gamma=56.8\pm0.5$\,km/s from de Smedt et al. (2016) as the systemic velocity 
for this star.

Second, the H$\alpha$ profile in the spectrum of V340\,Ser in Fig.\,\ref{Halpha-var} is a P\,Cyg type  one. 
This profile type is typical of high-luminosity stars with an extended
and expanding atmosphere. H$\alpha$ is formed in the upper layers of the extended stellar atmosphere.
The intense absorption component results from the absorption of radiation in the region in front 
of the star. The position of the core of the absorption
component (+34\,km/s) points to its formation in the high layers of the expanding atmosphere
close to the circumstellar medium. The velocity of about 34\,km/s differs from the adopted systemic
velocity, suggesting an expansion of the region in which this absorption is formed. Such a profile type
is rarely encountered in the spectra of post-AGB stars. Numerous examples of the varieties of the H$\alpha$
profile in the spectra of post-AGB stars are given by Klochkova~(1997), Sanchez Contreras et al.~(2008),
and Molina et al.~(2014): purely absorption profiles, complex P\,Cyg profiles, inverse P\,Cyg profiles, etc.

\begin{figure}[bpt!!!] 
\includegraphics[angle=0,width=0.8\textwidth, bb=30 40 705 520,clip]{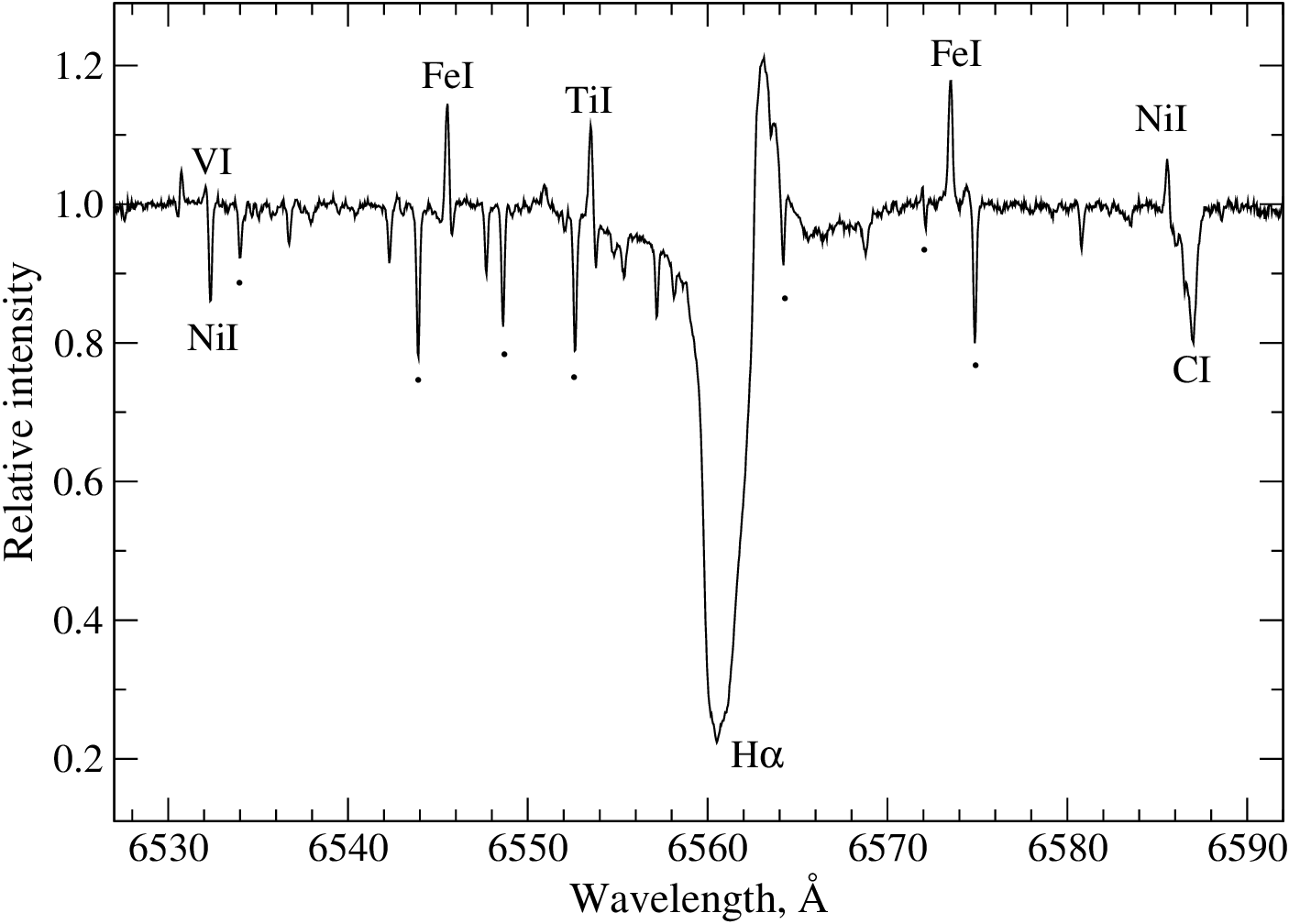}
\caption{Fragment of the spectrum for 89\,Her containing H$\alpha$ and TiI, VI, FeI, and NiI emissions. 
The identification of the main absorptions is indicated. The dots mark the telluric features.}
\label{89Her}
\end{figure}

However, in this extensive collection of recorded profiles there is no profile similar to 
the H$\alpha$ one in the spectra of V340\,Ser. A variety of profile types is observed in the 
papers by Gonzalez et al.~(1997) and Pollard et al.~(1997) devoted to the spectroscopy of
RV\,Tau stars, as in the community of post-AGB stars as a whole. An H$\alpha$ profile with  
a strong redshifted emission similar to the profile in the spectrum of V340\,Ser is encountered 
rarely, but, as has been noted above, it was recorded by Pollard et al.~(1997) in the spectrum 
of AD\,Aql. A significant phase variability of the H$\alpha$ profile, including the variability of
the positions of both its components, was recorded in the spectrum of this star. 
So far we have recorded no such significant variability of the positions and
intensities of the H$\alpha$ profile components for V340\,Ser. A change only in the extent 
of the blue absorption wing is clearly seen in Fig.\,\ref{Halpha-var}, tracing the 
variability of the outflow velocity. 
A close H$\alpha$ profile type with a large velocity  difference, $>$100\,km/s, for the absorption 
and emission components is also observed in the spectrum of the post-AGB star 89\,Her (=IRAS\,17534+2603). 
As an illustration, Fig.\,\ref{89Her}  shows a fragment of the spectrum for 89\,Her taken
on June 3, 2010, with the NES at the BTA telescope.

\begin{figure}[bpt!!!] 
\includegraphics[angle=0,width=0.5\textwidth, bb=30 40 705 530,clip]{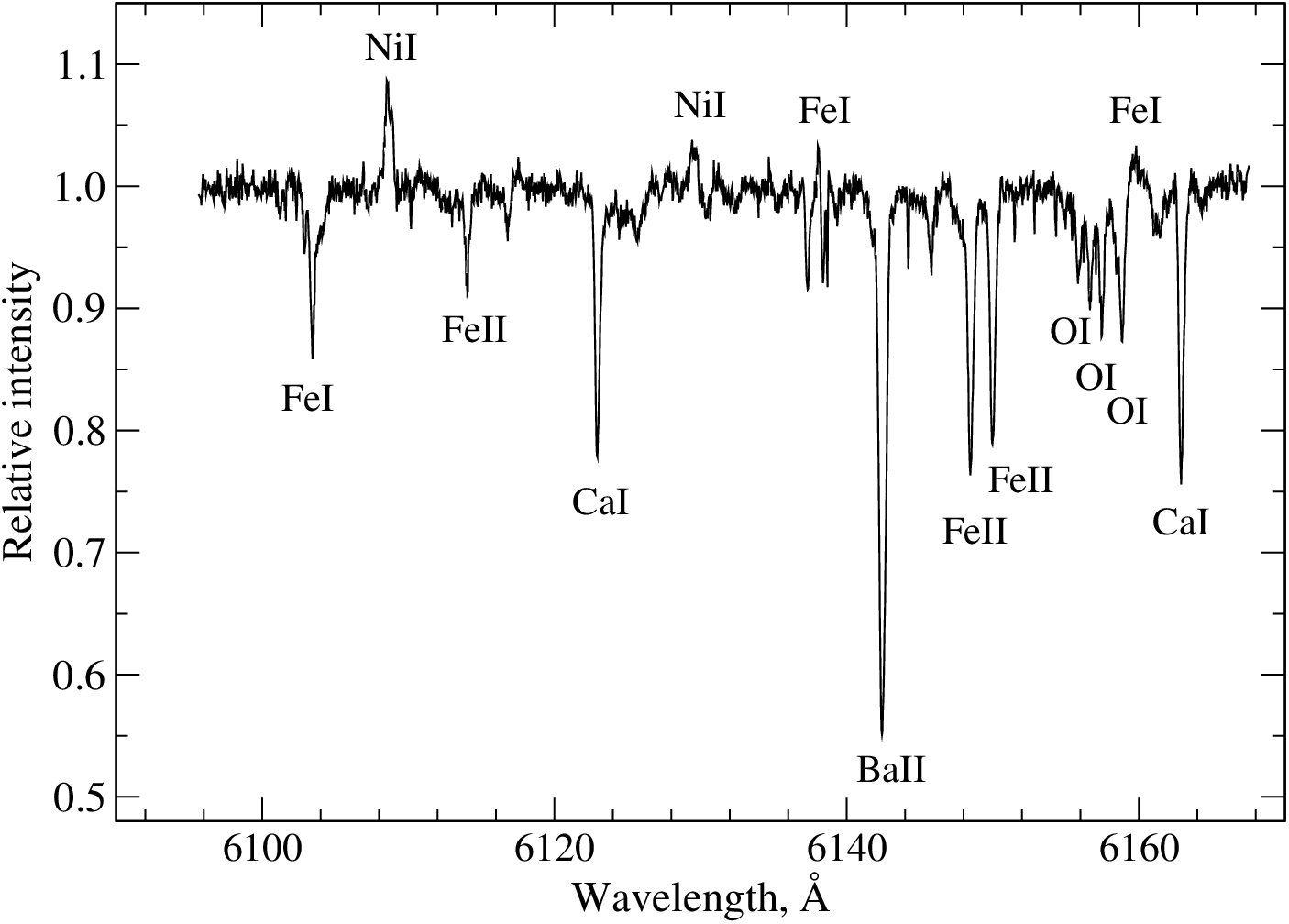}
\hspace{0.2cm}
\includegraphics[angle=0,width=0.5\textwidth, bb=30 40 705 530,clip]{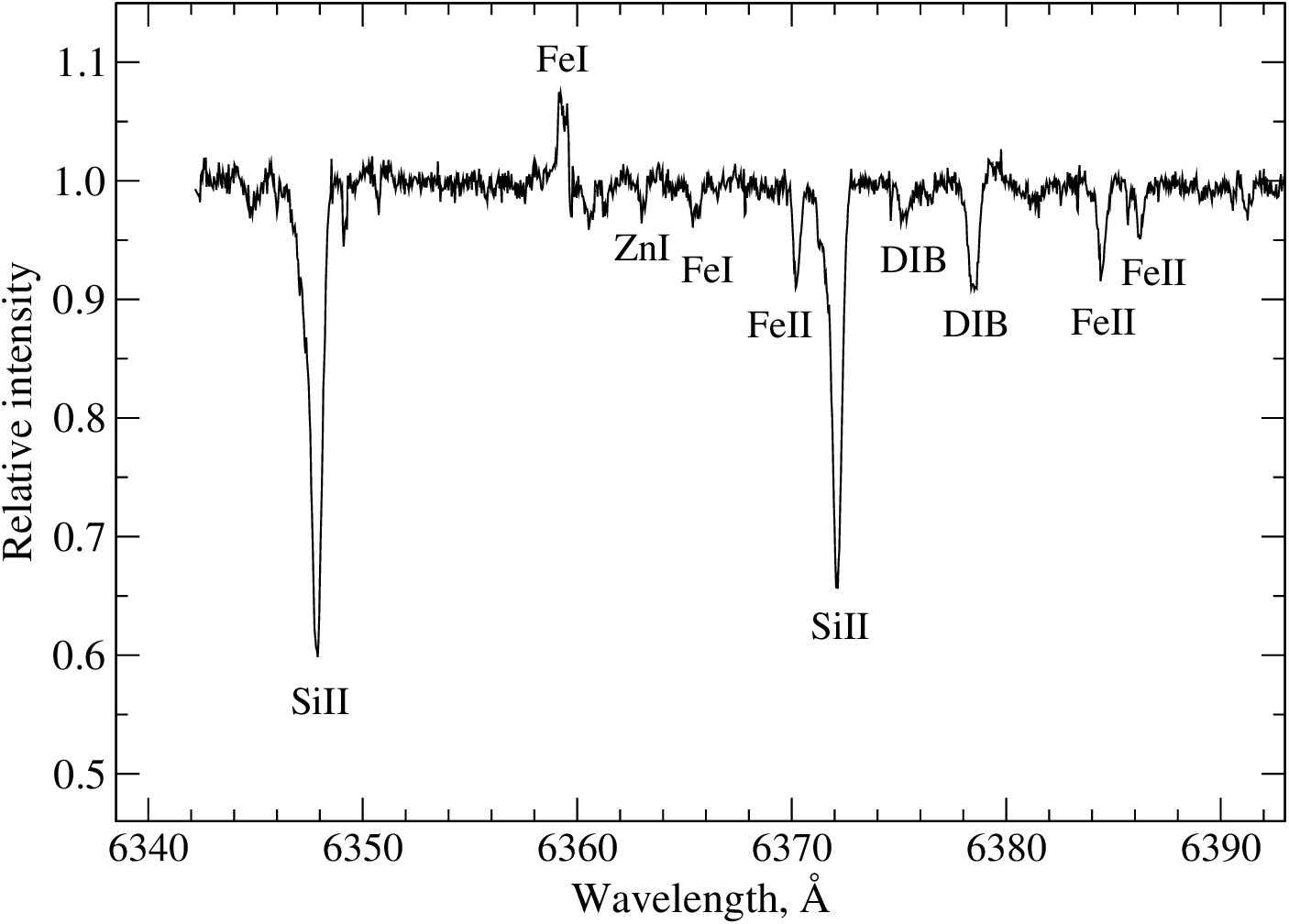}
\caption{Fragments of the spectrum for V340\,Ser containing peculiar features: FeI and NiI 
         emissions as well as strong SiII(2)\,6347 and 6371\,\AA{} absorptions with an extended blue 
         wing and DIBs in the right fragment. The identification of the  main features is indicated. }
\label{fragm}
\end{figure}

\begin{figure}[bpt!!!] 
\includegraphics[angle=0,width=0.5\textwidth, bb=10 70 550 740,clip]{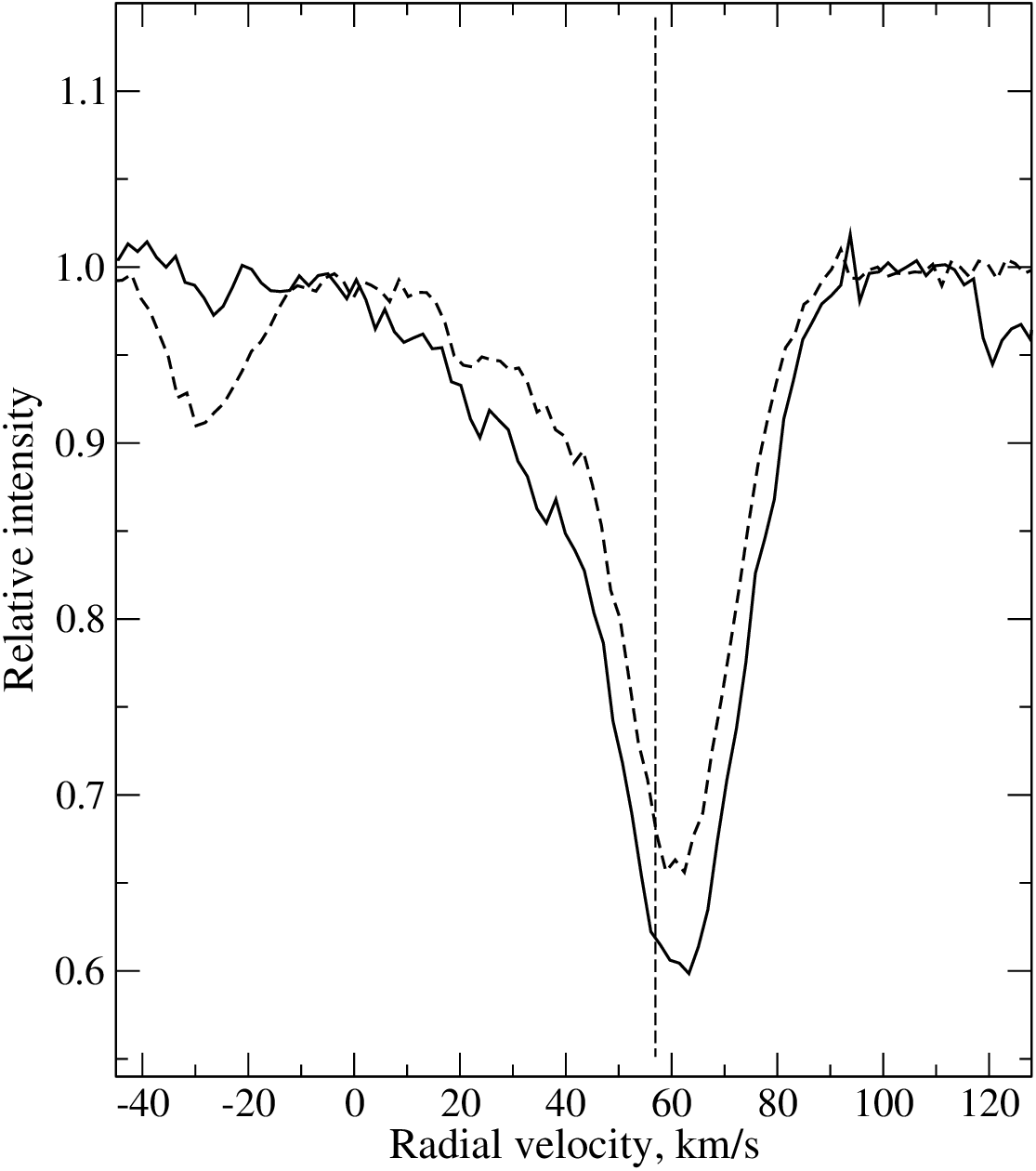}
\caption{SiII(2)\,6347 (solid line) and 6371\,\AA{} absorption profiles in the April 11, 2018 
       spectrum of V340 Ser. The vertical line marks the systemic velocity 
       $\gamma=56.8\pm0.5$\,km/s     (de Smedt et al., 2016). }
\label{Si}
\end{figure}

We see the influence of the wind distorting the blue wings of the strong SiII(2)\,6347 and 6371\,\AA{} 
absorptions in the spectrum of V340\,Ser in the right fragment of Fig.\,\ref{fragm}, where the FeI\,6359\,\AA{} 
emission and the $\lambda$\,=\,6376 and 6379\,\AA{} DIBs are also marked. 
The distortion of the SiII(2)\,6347 and 6371\,\AA{}  absorption wings in the profiles in relative 
intensity -- radial velocity coordinates is also illustrated by Fig.\,\ref{Si}.
Here the absorption cores of both proﬁles are shifted relative to the adopted systemic velocity, 
but their positions agrees well with the velocity from different metal absorptions in the spectrum: 
in Table\,\ref{velocity} for this date V$_{\sun}$(abs)\,=\,60.09\,km/s.

Having obtained the set of radial velocities for V340\,Ser by the cross-correlation method, de Smedt
et al.~(2016) concluded that the star is binary and classified V340\,Ser as a spectroscopic binary with a
period of 365~days. Later, using the same technique, Oomen et al.~(2018) improved the orbital elements for
a sample of post-AGB stars in binary systems. For V340\,Ser they provide the following orbital elements:
a period of 363 days, a $\gamma$ velocity of 56.1\,km/s, and a velocity amplitude K\,=8.4$\pm$1.0\,km/s. 

Note two points. First, having high-quality spectra of this star, we found no evidence of spectroscopic binarity.
Second, as follows from the data in Table\,\ref{velocity}, having taken the spectra of V340\,Ser at arbitrary 
dates in 2018--2020, we recorded no radial velocity variability in the star expected from the results by de Smedt
et al.~(2016) and Oomen et al.~(2018). Given the period about 1 year, we can assume that all our spectra
were taken at close orbital phases. The necessity of continuing the spectral observations of V340\,Ser 
and a detailed study of the variability of the velocity pattern is obvious.

\subsection{Metal emissions}

The spectral fragments in Fig.\,\ref{fragm}  illustrate the following peculiarity of the optical 
spectrum for V340 Ser that has not been noted previously in publications: the presence of weak 
emissions of neutral metals with a low excitation potential of the lower level. Two such 
emissions in the H$\alpha$ wings are also clearly seen in Fig.\,\ref{Halpha-var}. 
Table\,\ref{emis} lists all of such emissions identified by us in the recorded
wavelength range. The last column in this table gives the velocities corresponding to the
position of the emission feature. As follows from the data in Table\,\ref{velocity}, the mean 
velocity from the emissions changes insignificantly from date to date, from 
V$_{\sun}$(em)\,=45.0$\pm$0.2 to 48.0$\pm$0.4\,km/s. For each date V$_{\sun}$(em)  differs 
systematically by $\ge10$\,km/s from the mean velocity inferred from the absorptions,
suggesting a velocity stratification in the extended stellar atmosphere. The full widths 
at half maximum (FWHMs) of these emissions forming in the outer layers of the extended 
atmosphere of V340\,Ser are about 0.25\,\AA{} or $\delta$V$_{\sun}\approx 10\div12$\,km/s, 
which is twice the FWHM of the forbidden ionospheric [OI] emissions in the spectrum.
It should be emphasized that emissions are also present in several ion lines, for example, 
NdII\,5102, YII\,5509  and EuII\,6645\,\AA{}.  This peculiarity can affect the abundances 
of chemical elements and, therefore, it should be confirmed in subsequent observations.

Such low-excitation emissions of neutral metals were previously recorded in the spectra of 
the post-AGB candidates LN\,Hya (=IRAS\,12538$-$2611) (Klochkova and Panchuk, 2012) and V1648\,Aql
(Klochkova and Tavolzhanskaya, 2019), whose spectral types are close to that of V340\,Ser. 
For LN\,Hya metal emissions appeared in the spectra taken at its active phases in 2010, when 
the inverse P\,Cyg profile for the metal lines differed significantly from the one
observed at quiescent phases. The position of the H$\alpha$ absorption component also 
differed significantly from its position at other dates of observations. Furthermore, 
the H$\alpha$ core was significantly (approximately by 15\,km/s) redshifted relative to 
the symmetric metal absorptions. 

The spectrum of the above mentioned post-AGB star 89\,Her  also contains a lot of low excitation 
emissions, as was already noted previously by Waters et al.~(1993). As an illustration,
Fig.\,\ref{89Her} presents a fragment with narrow emissions of the spectrum for this star 
taken with the NES. Emissions of metal atoms were also recorded in the 1980-s in the spectra 
of RV\,Tau stars. We will also add the paper by Bopp~(1984), whose author emphasized the 
appearance of such emissions during deep minima of the star R\,Sct, which also belongs to 
RV\,Tau stars, to the already mentioned papers by Baird~(1982, 1984) and Klochkova and
Panchuk~(1998).

Some of the Fe, Co, and Ni emissions from Table\,\ref{emis} are also presented in the spectrum 
of the yellow hypergiant $\rho$\,Cas with an extended envelope (for details and references, 
see Klochkova et al.~2018). Besides, in the spectrum of $\rho$\,Cas the mean velocity from these
emissions changes insignificantly with time and differs little from the systemic velocity of the 
hypergiant. The low width of these emissions in the spectrum of $\rho$\,Cas and the coincidence 
of the velocities suggest that these weak emissions are formed in the outer extended gaseous 
envelope whose sizes exceed considerably the photometric radius of the star. Emissions are 
observed predominantly in the periods of brightness decline in $\rho$\,Cas, which may point to a
relative stability of the emission measure observed against the background of a weakened photospheric
spectrum.

\begin{longtable}{l llc@{\quad}}
\caption{Weak emissions identified in the spectrum of V340\,Ser}      
\\ \hline
\endfirsthead
\hline
\multicolumn{3}{l}{Table\,2, cont} \\ \hline
\endhead
\hline
\multicolumn{3}{r}{\it see next page } \\ \hline
\endfoot
\hline
\endlastfoot
$\lambda$  & Element &  V$_{\sun}$, km/s \\ 
\hline 
4953.200  & NiI  &  46.20    \\   
5079.682  & CeII &  45.86    \\   
5092.800  & NdII &  42.86    \\   
5102.390  & NdII &  42.27    \\   
5107.446  & FeI  &  45.85    \\   
5150.840  & FeI  &  43.08    \\   
5151.910  & FeI  &  45.20    \\   
5198.710  & FeI  &  43.00    \\   
5216.270  & FeI  &  43.39    \\   
5409.784  & CrI  &  42.29    \\   
5434.523  & FeI  &  40.45    \\   
5462.485  & NiI  &  41.89    \\   
5490.150  & TiI  &  44.67    \\   
5644.140  & TiI  &  48.95    \\   
5711.883  & NiI  &  46.81    \\   
5847.000  & NiI  &  47.13    \\   
5999.040  & TiI  &  42.84    \\   
6007.310  & NiI  &  43.97    \\   
6065.482  & FeI  &  44.89    \\   
6108.107  & NiI  &  48.32    \\   
6108.530  & CI   &  40.30    \\   
6128.990  & NiI  &  42.83    \\   
6137.691  & FeI  &  43.60    \\   
6191.190  & NiI  &  46.14    \\   
6191.558  & FeI  &  45.09    \\   
6200.313  & FeI  &  43.19    \\   
6213.429  & FeI  &  43.55    \\   
6219.270  & FeI  &  44.89    \\   
6230.722  & FeI  &  42.47    \\   
6261.098  & TiI  &  43.08    \\   
6265.131  & FeI  &  45.17    \\   
6312.240  & TiI  &  43.95    \\   
6330.130  & CrI  &  45.75   \\    
6358.693  & FeI  &  48.43   \\    
6393.600  & FeI  &  43.43   \\    
6400.000  & FeI  &  43.00   \\    
6430.844  & FeI  &  44.75   \\    
6494.980  & FeI  &  42.93   \\    
6498.950  & FeI  &  48.96   \\    
6531.410  & VI   &  47.03   \\    
6532.890  & FeI  &  41.82   \\    
6546.238  & FeI  &  45.25   \\    
6554.230  & TiI  &  46.92   \\    
6572.800  & CaI  &  45.91   \\    
6574.240  & FeI  &  46.57   \\    
6592.913  & FeI  &  43.03   \\    
6643.629  & NiI  &  46.40   \\    
6677.985  & FeI  &  43.14   \\    
6767.768  & NiI  &  43.86   \\    
6770.960  & CoI  &  46.38   \\    
6814.950  & CoI  &  47.39   \\    
7052.870  & CoI  &  46.81   \\    
7138.910  & TiI  &  46.60   \\    
7291.449  & NiI  &  41.72   \\  
7357.740  & TiI  &  47.17   \\      
7714.308  & NiI  &  48.73   \\
\hline
\label{emis}
\end{longtable}

\subsection{DIBs and the multicomponent profile of the NaI~D lines}

Despite its significant distance from the Galactic plane (the Galactic latitude of the star $|b|>12\degr$),
numerous interstellar features were detected in the optical spectrum of V340\,Ser. In particular, two
such features with wavelengths $\lambda$\,=\,6376 and 6379\,\AA{} are clearly seen in Fig.\,\ref{fragm}. 
Table\,\ref{DIBs} lists all of the DIBs from the well-known list by Jenniskens and Desert~(1994) 
identified by us and reliably distinguished among the blends in the spectrum of V340\,Ser. 
For these features the table gives the radial velocity corresponding to the DIBs positions and their
equivalent widths W$_{\lambda}$.

\begin{figure}[h!] 
\includegraphics[angle=0,width=0.5\textwidth, bb=10 70 550 720,clip]{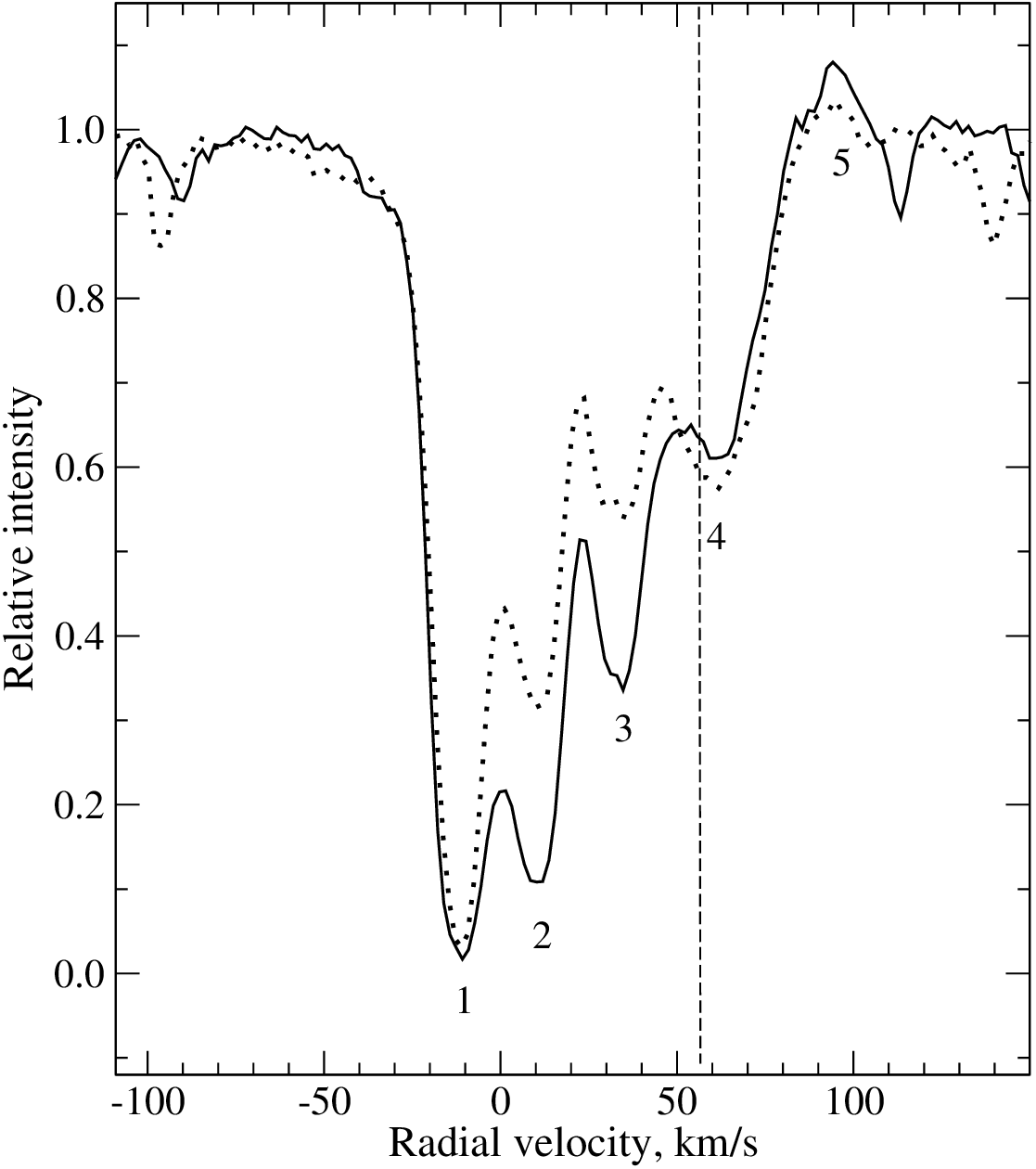}
\caption{NaI\,5889 and 5895\,\AA{} line profiles (dots) in the spectrum of V340\,Ser. 
     The positions of the profile absorption components forming in the interstellar medium (1), 
     the circumstellar medium (2 and 3), and the stellar atmosphere (4) are marked. The
     vertical line marks the systemic velocity  $\gamma=56.8\pm0.5$\,km/s (de Smedt et al., 2016). }
\label{NaD}
\end{figure}

A high quality of our spectra for V340\,Ser allowed us to resolve  the NaI\,5889 and 5895\,\AA{}~D 
lines into components and to measure the position of the interstellar  KI~7696\,\AA{} absorption. 
Individual absorptions, whose average positions are given in Table\,\ref{velocity} and
Fig.\,\ref{NaD}, are confidently distinguished in the NaI\,5889 and 5895\,\AA{} lines profile in 
our spectra. The position of the long-wavelength absorption component (designated as 4 in this figure) 
for each date of observations agrees, within the above error limits, with the mean velocity 
V$_{\sun}$(abs) measured from a large set of metal absorptions. Such agreement suggests the formation
of this NaI~D line component in the stellar atmosphere. 

According to the data from Table\,\ref{velocity}, the position of the shortest-wavelength component~1 
of the NaI doublet lines, V$_{\sun}=-$11.3\,km\,s, is stable. Given the measurement errors, its 
position coincides with the average velocity from the set of DIBs identified in
the spectrum, suggesting that this NaI doublet component is formed in the interstellar medium. 
In Fig.\,\ref{NaD} we can see a differing steepness of the component wings, which also confirms 
the proposed interpretation of their formation regions.

\begin{table*}[h]
\medskip
\caption{Parameters of DIBs in the spectrum of  V340\,Ser}
\begin{tabular}{ c|  c  r }
\hline
 $\lambda$, \AA{}&\hspace{3mm} V$_{\sun}$, km/s & \hspace{2mm} W$_{\lambda}$, m\AA{} \\  
 \hline 
4984.79 &   $-11.86$  &  23  \\ 
4987.42 &   $-12.70$  &  6   \\ 
5780.48 &   $-12.55$  &  263  \\
5797.06 &   $-12.11$  &  49 \\  
5818.75 &   $-9.61$   &  1 \\   
5849.81 &   $-10.69$  &  5  \\  
6195.98 &   $-11.35$  &  36  \\ 
6203.05 &   $-10.18$  &  52  \\ 
6211.60 &   $-11.39$  &  29  \\ 
6269.85 &   $-11.04$  &  12  \\ 
6283.84 &   $-10.27$  &  85:\\  
6329.97 &   $-12.75$  &  7 \\   
6376.08 &   $-12.41$  & 18  \\  
6379.32 &   $-10.40$  &  41 \\  
6439.48 &   $-11.48$  &  4 \\   
6445.28 &   $-13.03$  &  7 \\   
6532.10 &   $-12.14$  & 11 \\   
6613.62 &   $-12.85$  & 114 \\  
6632.86 &   $-11.52$  & 9 \\    
6770.05 &   $-10.20$  & 7 \\    
6827.30 &   $-11.26$  & 7 \\    
7224.03 &   $-10.76$  & 9 \\    
7357.20 &   $-10.48$  &17 \\    
7367.13 &   $-13.50$  &13  \\   
7581.30 &   $-10.22$  & 3  \\   
7721.85 &   $-9.80$   & 4   \\  
\hline                
\end{tabular}   
\label{DIBs}
\end{table*}

Components 2 and 3 of the NaI~D lines with velocities V$_{\sun}$=+10.1 and +33.7\,km/s are redshifted
relative to the mean radial velocity V$_{\sun}$(abs) from the photospheric absorptions. It is natural 
to assume that both these components are formed in the circumstellar medium and the outer layers 
of the expanding stellar atmosphere. The component with a velocity of 10.1\,km/s is formed in an envelope 
expanding with a velocity typical of the envelopes of post-AGB stars (for comparison, see the results by 
Bakker et al.~(1995, 1997) and Klochkova~(2014)). The component with a velocity of 33.7\,km/s is formed in the
uppermost layers of the stellar atmosphere that pass into the circumstellar medium undetached from the
star. The absorption of the complex H$\alpha$ profile is also formed here. These layers move with 
a velocity $\approx$23\,km/s relative to the deep atmospheric layers, where symmetric absorptions are 
formed. The multicomponent profile of the NaI doublet lines also contains the emission component 5 
exceeding the local continuum in Fig.\,\ref{NaD}. The position of this emission,
V$_{\sun}\approx 104$\,km/s, is close to the position of the H$\alpha$ emission component.

Note the intensity ratio of the components in the NaI~D1 (5895\,\AA{})  and D2 (5889\,\AA{}) lines.  
If the corresponding NaI atomic levels are populated proportionally to the statistical weights, 
then the D2/D1 ratio in the emission spectrum must be 2.  In the spectra of V340\,Ser the emission 
feature 5 is clearly seen in the D2 line and is at the detection limit in the D1 line. As the 
number density of NaI  atoms rises, the role of the resonant scattering and self-absorption 
processes increases and, as a result, the intensity ratio of the NaI emission doublet lines
will tend to 1. 

In the spectra of stellar atmospheres we observe the D1 and D2 absorptions of equal intensities 
(for example, component 4 in Fig.\,\ref{NaD}). As a rule, a similar picture is also observed in 
the spectra of the interstellar medium (for example, component 1
in Fig.\,\ref{NaD}), suggesting a saturation of the NaI doublet absorption profiles 4 and 1 
due to self-absorption. In the spectra of the circumstellar envelopes of V340\,Ser
the intensity ratios of the doublet absorption components differs significantly: $\approx$1.3 and 
1.5 for components 2 and 3, respectively. We conclude that no saturation in the line cores has 
been achieved for both emission feature 5 and absorption features 2 and 3.

Such a multicomponent profile of the NaI~D lines with a fragment of the long-wavelength 
emission was recorded previously by Klochkova and Chentsov~(2004) in the spectrum of the post-AGB
star V510\,Pup (=IRAS 08005$-$2356). This star has evolved farther from the AGB than V340\,Ser
and, therefore, has a detached envelope, which manifests itself, in particular, in the presence of
circumstellar Swan~C$_2$ absorption bands. Furthermore, emission from the OH bands simultaneously 
at 1612 and 1667\,MHz was recorded for the source IRAS\,08005$-$2356, which, according to
Lewis~(1989), in the absence of SiO and H$_2$O masers suggests that the object approaches 
the planetary nebula stage. Note that in the spectrum of V510\,Pup Klochkova and Chentsov~(2004) 
identified the forbidden [CaII]\,7291 and 7324\,\AA{} doublet emissions rarely observed in the 
spectra of post-AGB stars.
These low excitation forbidden emissions are typical for the spectra of stars of selected types with a mass
outflow and serve as evidence for the presence of a circumstellar disk. In particular, these emissions
were identified in the spectra of the yellow supergiants V1302\,Aql (=IRC+10420) (Klochkova et al.~1999)
and V509\,Cas (=IRC+60379) (Aret et al.~2017; Klochkova et al.~2019; Klochkova~2019) and the spectra of 
B supergiants with the B[e] phenomenon (Aret et al. 2012). In the spectra of V340\,Ser we
also see an emission that could be identified with the forbidden 7291\,\AA{} line, which could point to
the probable presence of a circumstellar disk in the system. Unfortunately, the second line  of this doublet, 
[CaII]\,7324\,\AA{}, which could confirm the presence of a forbidden emission, falls between adjacent 
orders in our echelle frame. Additional observations are needed to confirm the presence of this important 
spectral feature. At present, we have to identify the 7291\,\AA{}  emission in the spectrum with the 
low xcitation NiI line with $\lambda$=7291.449\,\AA{}.

Emission components in the NaI resonance lines with a FWHM $\Delta \approx 120$\,km/s were recorded by
Klochkova et al.~(2006) in the spectrum of the cool variable AGB star associated with the IR source
IRAS\,20508+2011. The multicomponent profile of the NaI~D lines in the spectrum of the semiregular 
variable QY\,Sge is even more interesting. A broad emission with a FWHM $\Delta \approx 120$\,km/s 
was recorded by Menzies and Whitelock~(1988), who assumed that the formation region of the emission in
the resonance lines is a fairly hot and inhomogeneous circumstellar envelope, while the large emission
width is due to the scattering of photons by moving dust grains in the envelope. Having recorded both
broad and narrow metal emissions in the spectrum of QY\,Sge, Rao et al.~(2002) developed the assumption
of Menzies and Whitelock~(1988) by proposing a model with a circumstellar torus and bipolar jets.
Subsequently, based on their spectroscopic monitoring of QY\,Sge, Klochkova et al. (2007b) confirmed
the conclusion about the emission formation in the circumstellar medium by showing the stability of its
position in time.

\subsection{Distance to the star  and its luminosity}

We will estimate the luminosity of the star using the luminosity -- OI\,7774\,\AA{} triplet intensity 
criterion known for F-G supergiants. In the spectra of V340\,Ser the total equivalent width of the 
oxygen triplet is W$_{\lambda}$(7774)\,=\,1.25\,\AA{},  which is a typical value for post-AGB stars 
(for comparison, see the data from Molina et al.~(2018)). The luminosity estimate for V340 Ser from  
W$_{\lambda}$(7774)  is fairly accurate, because, according to van Winckel~(1997), there is no oxygen 
abundance anomaly in the atmosphere of this star. Applying the calibration 
Mv$\longleftrightarrow \rm{W_{\lambda}}$(7774) from Takeda et al.~(2018), we get an absolute magnitude 
Mv$\approx -4.6^m$ 
for V340\,Ser. The luminosity of V340\,Ser derived in this way, log$L/L_{\sun}$=\,3.75, agrees well with 
its spectral classification and the luminosity expected from the theoretical views of Bl\"ocker
(1995b) about the evolutionary status of post-AGB stars.

To estimate the color excess $E(B-V)$ due to the interstellar extinction, we used the equivalent widths
W$_{\lambda}$(DIBs) measured in the spectra of V340\,Ser for a sample of DIBs and the calibration 
relations $E(B-V)\longleftrightarrow \rm{W}_{\lambda}$(DIBs) from Kos and Zwitter (2013). 
Invoking seven confidently measured bands from Table\,\ref{DIBs} (4984, 5780, 5797, 6196, 6203, 6379, 6613\,\AA{}),
for which the calibration relations by Kos and Zwitter~(2013) are available, we found the mean color
excess to be $E(B-V)=0.59^m$. This color excess is definitely below the total reddening 
$E(B-V)$, because some fraction of the stellar reddening is also attributable to the extinction 
in the circumstellar medium. As follows from Fig.\,\ref{NaD}, the intensity of the circumstellar 
components 2 and 3 is only slightly lower than that of the interstellar one, suggesting a
significant contribution of the circumstellar extinction to the total reddening $E(B-V)$. 

By modeling the spectral energy distribution (SED) for V340\,Ser, de Smedt et al.~(2016) estimated 
the total reddening: $E(B-V)=0.76^m$. With this estimate of the color excess we find the total 
extinction to be Av$\approx 2.43^m$ (for the standard ratio  R\,=\,3.2). Taking into account the mean 
apparent magnitude V\,=\,9.63$^m$ from Arkhipova et al.~(2011), we estimate the distance to the star 
to be d$\approx$2.3\,kpc. The distance estimate d$\approx$3.43\,kpc for V340\,Ser from the extensive
catalogue by Vickers et al.~(2015) should be mentioned here. While modeling the SED for post-AGB
stars, these authors adopted the luminosity range $L/L_{\sun}=6000\pm1500$ for them. Our luminosity 
estimate for V340\,Ser from the IR oxygen triplet lies within this range. While modeling the SED, 
Vickers et al.~(2015) took into account the anomalous SED pattern for IRAS\,17279$-$1119 due to 
the presence of warm ($\approx$855\,K) and cold ($\approx$171\,K) dust. At the same time, Vickers 
et al.~(2015) found the reddening to be $E(B-V)$=0.45$^m$, which is considerably lower than
the reddening from de Smedt et al.~(2016) and even lower than the interstellar reddening fraction 
derived by us from DIBs.

The Gaia DR2 catalogue~(2018) gives a reliable parallax for V340\,Ser, $\pi$=0.1577$\pm$0.0427\,mas,
which leads to a significant distance of the star: d$\ge$6.3\,kpc. However, in the 2016 version the data for
the star differ significantly: $\pi$=0.340$\pm$0.260\,mas. This parallax gives a distance to the star 
close to our estimate, but the parallax accuracy is too low. Here we should allude to the paper by 
Xu et al.~(2019), in which they compared the Gaia DR2 parallaxes with the VLBI data and noted a large 
difference in these data for AGB stars with extended dust envelopes. Furthermore, the large parallax 
offset for Gaia DR2 relative to VLBI, $-0.075\pm$0.029\,mas (Xu et al.~2019), should be taken into 
account for distant objects.

On the whole, it can be asserted that the set of parameters for V340\,Ser, i.e., its luminosity, distance
from the Galactic plane, lowered  metallicity, and chemical peculiarities, is consistent with its belonging 
to the post-AGB stage in the Galactic thick disk. Note that the set of main parameters and chemical composition
of V340\,Ser allows this star to be included in the homogeneous group of post-AGB stars with large
enhancements of carbon and heavy s-process metals identified by Klochkova (1997) and van Winckel and
Reyniers~(2000). All stars of this group have structured,  often bipolar, envelopes, where a lot of 
molecular and maser features are formed. Furthermore, an as yet unidentified emission at 21$\mu$ is 
contained in the IR spectra of all these stars (for more details, see Hrivnak et al.~2009). 
The circumstellar medium of V340\,Ser has no these features, i.e., its envelope has not yet detached 
from the stars and has not cooled to the requires extent. 
Hrivnak et al. (2009) suggest that the feature at 21$\mu$ is formed in the IR spectra
of post-AGB stars with cold dust envelopes, with a temperature $\approx$120--150\,K.

It is necessary to check whether V340 Ser belongs to RV\,Tau stars,  because a star with the main
parameters Teff=\,7200\,K and  $\log L/L_{\sun}$=3.75  lies outside the instability strip, according to the data by
Kiss et al.~(2007). Recall that, previously, Arkhipova et al.~(2011) also questioned this status 
of V340\,Ser, while Si\`odmiak et al.~(2008) classified V340\,Ser as a post-AGB star, but not as 
an RV\,Tau one.

\section{CONCLUSIONS}

We studied the optical  spectra of V340\,Ser (=IRAS 17279$-$1119) taken at the 6-m BTA telescope with
a spectral resolution R$\ge$60000. For four dates we measured the heliocentric radial velocity from
numerous metal absorptions (from 300 to 550 lines):
V$_{\sun}$=\,59.30$\pm$0.05, 60.09$\pm$0.05, 58.46$\pm$0.06, and 55.78$\pm$0.06\,km/s. Given the high 
accuracy of the velocity, these values of  point to a weak variability
of the velocity with time.

The interstellar,  V$_{\sun}$=$-11.20$\,km/s, and circumstellar, V$_{\sun}$=+10\,km/s, 
components and the component forming in the uppermost atmospheric layers, V$_{\sun}$=+34.0\,km/s, 
were identified in the NaI~D line profile. The values averaged over four spectra have an accuracy of 
$\pm$0.2\,km/s. The mean velocity from 20--30~DIBs identified in the spectra,
 V$_{\sun}$(DIBs)=$-11.6\pm$0.2\,km/s, coincides with the velocity from the interstellar NaI component. 
 
The H$\alpha$ line has a P\,Cyg type profile, with the position of its absorption component (V$_{\sun}$=+34\,km/s) 
pointing to its formation in the upper layers of the expanding atmosphere.

The weak emissions with an intensity $\approx$10\% of the local continuum level were identified 
with low excitation lines of metal atoms. Their stable positions, V$_{\sun}$=46.3$\pm$0.4\,km/s, 
systematically differ from the velocity inferred from atmospheric absorptions, suggesting the 
presence of a velocity gradient in the upper layers of the stellar atmosphere.

Based on the equivalent width triplet  W$_{\lambda}$(7774)=1.25\,\AA{} of the IR oxygen, we estimated 
the absolute magnitude to be Mv$\approx-4.6^m$, which, when taking into account  the total (in the interstellar and 
circumstellar media)  extinction, leads to a distance estimate for the star d$\approx$2.3\,kpc. 
This value is half the distance corresponding to the parallax of V340\,Ser from DR2 Gaia~(2018). 
Such a difference can be a consequence of the correction for the extinction in the interstellar 
medium and the object’s circumstellar environment. 

It is necessary to check whether V340\,Ser belongs to RV\,Tau type stars, because a star with the main
parameters Teff\,=\,7200\,K and  $\log L/L_{\sun}$=\,3.75  lies outside the instability strip.

\acknowledgement We thank the Russian Foundation for Basic Research for its partial financial support 
(project no.\,18-02-00029 a). We used the SIMBAD, SAO/NASA ADS, ASAS, and Gaia DR2 
astronomical databases.

\newpage

\section*{References}

\begin{itemize} 

\item A. Arellano Ferro, S. Giridhar, and P. Mathias, Astron. Astrophys.  368, 250 (2001).

\item  F. Arenou, X. Luri, C. Babusiaux, et al., Astron. Astrophys. 616, A17 (2018); 

\item  A. Aret, M. Kraus, M. F. Muratore, and M. Borges Fernandes, MNRAS, 423, 284 (2012).

\item  A. Aret, M. Kraus, I. Kolka, and G. Maravelias, ASP Conf. Ser. 508, 357 (2017).
 
\item  V.P. Arkhipova, N. P. Ikonnikova, and G. V. Komissarova, Astron. Lett. 37, 635 (2011).
 
\item  S.R. Baird, PASP, 94, 850 (1982).

\item  S.R. Baird, PASP 96, 72 (1984).

\item  8.E. Bakker, H. J. G. L. M. Lamers, L. B. F. M. Waters, and T. Schoenmaker, Astrophys. Space Sci. 224, 335
   (1995).
   
\item    E. Bakker, E. F. van Dishoeck, L. B. F. M. Waters, and T. Schoenmaker, Astron. Astrophys. 323, 469 (1997).

\item   T. Bl\"ocker, Astron. Astrophys. 297, 727 (1995a).

 \item  T. Bl\"ocker, Astron. Astrophys. 299, 755 (1995b).

 \item  B.W. Bopp, PASP, 96, 432 (1984).

 \item  M. di Criscienzo, P. Ventura, D. A. Garcia Hernandez, et al.  
             MNRAS,  462, 395 (2016).

\item   G.A. Galazutdinov, SAO Preprint No. 92 (1992).

\item  S. Giridhar, D. L. Lambert, and G. Gonzalez, Astrophys. J. 531, 521 (2000).

\item   G. Gonzalez, D. L. Lambert, and S. Giridhar, Astrophys. J. 479, 427 (1997).

\item  F. Herwig, Ann. Rev. Astron. Astrophys. 43, 435 (2005).
  
\item  B.J. Hrivnak, K. Volk, and S. Kwok, Astrophys. J. 694, 1147 (2009).

\item   P. Jenniskens and F.-X. Desert, Astron. Astrophys. Suppl. Ser. 106, 39 (1994).

\item   N. Rao Kameswara, A. Goswami, and D. L. Lambert, MNRAS,  334, 129 (2002)

\item   E.V. Kazarovets, N. N. Samus, and O. V. Durlevich, Inform. Bull. Var. Stars, No. 4870, 1 (2000).

\item   T. Kipper and V. G. Klochkova, Baltic Astron. 22, 77 (2013).

 \item   L.L. Kiss, A. Derekas, Gy. M. Szabo,  T. R. Bedding, and L. Szabados, MNRAS. 375,
1338 (2007).

\item   V.G. Klochkova, MNRAS, 272, 710  (1995).

 \item  V.G. Klochkova, Bull. Spec. Astrophys. Observ. 44, 5 (1997).

 \item   V.G. Klochkova, Astrophys. Bull. 69, 279 (2014).

 \item V.G. Klochkova, Astrophys. Bull. 74, 475 (2019).

\item  V.G. Klochkova and V. E. Panchuk, Astron. Lett. 24, 650 (1998).

\item  V.G. Klochkova and V. E. Panchuk, Astron. Rep. 56, 104 (2012).

 \item V.G. Klochkova and N. S. Tavolzhanskaya, Astrophys. Bull. 74, 277 (2019).
 
\item  V.G. Klochkova and E. L. Chentsov, Astron. Rep. 48, 301 (2004).
 
\item  V.G. Klochkova, E. L. Chentsov, and N. S. Tavolganskaya, Astrophys. Bull. 48, 25 (1999).

\item  V.G. Klochkova, V. E. Panchuk, N. S. Tavolzhanskaya, and G. Zhao, Astron. Rep. 50, 232 (2006).

\item V.G. Klochkova, E. L. Chentsov, N. S. Tavolganskaya, and M. V. Shapovalov, Astrophys. Bull. 62, 162 (2007a).

\item  V.G. Klochkova, V. E. Panchuk, E. L. Chentsov, and M. V. Yushkin, Astrophys. Bull. 62, 162 (2007b).

\item  V.G. Klochkova, V. E. Panchuk, and N. S. Tavolzhanskaya, Astron. Rep. 62, 623 (2018).

\item   V.G. Klochkova, E. L. Chentsov, and V. E. Panchuk, Astrophys. Bull. 74, 41 (2019).

\item   J. Kos and T. Zwitter, Astrophys. J. 774, 72 (2013).

\item  B.M. Lewis, Astrophys. J. 338, 234 (1989).

\item  N. Liu, R. Gallino, S. Bisterzo, A. M. Davis, R. Trappitisch, and L. R. Nittler, Astrophys. J. 865, 112 (2018).

\item J.W. Menzies and P. A. Whitelock, MNRAS,  233, 697 (1988).

\item  R.E. Molina, S. Giridhar, C. B. Pereira, A. Arellano Ferro, and S. Muneer, Rev. Mex. Astron. Astroﬁs. 50, 293 (2014).

\item  G.-M. Oomen, H. van Winckel, O. Pols, et al. , Astron. Astrophys. 620, A85 (2018).

\item  R.D. Oudmaijer, W. E. C. J. van der Veen, L. B. F. M. Waters, et al. 
        Astron. Astrophys. Suppl. Ser. 96, 625 (1992).

\item  V.E. Panchuk, V. G. Klochkova, and M. V. Yushkin, Astron. Rep. 61, 820 (2017).

\item   K.R. Pollard, P. L. Cottrell, W. A. Lawson, M. D. Albow, and W. Tobin, MNRAS,  286, 1 (1997).

\item S. Sumangala Rao, S. Giridhar, and D. L. Lambert, MNRAS, 419, 1254 (2012).

\item  C. Sanchez Contreras, R. Sahai, R. Goodrich, and A. Gil de Paz, Astrophys. J. Suppl. Ser. 179, 166
     (2008).

\item  N. Si\`odmiak, M. Meixner, T. Ueta,   et al. 
         Astrophys. J. 677, 382 (2008).

\item  K. de Smedt, H. van Winckel, D. Kamath,   et al.   
Astron.    Astrophys. 587, A6 (2016).

\item   Y. Takeda, G. Jeong, and I. Han, Publ. Astron. Soc. Jpn. 70, 8 (2018).

\item  S.B. Vickers, D. J. Frew, O. A. Parker, and I. S. Bojicic, MNRAS,  447, 1673 (2015).

\item  L.B.F.M. Waters, C. Waelkens, M. Mayor, and N. R. Trams, Astron. Astrophys. 269, 242 (1993).

\item H. van Winckel, Astron. Astrophys. 319, 561 (1997).
 
\item  H. van Winckel and M. Reyniers, Astron. Astrophys. 354, 135 (2000).

\item   S. Xu, B. Zhang, M. J. Reid, Z. Xingwu, and W. Guangli, Astrophys. J. 875, 114 (2019).

\item   M.V. Yushkin and V. G. Klochkova, SAO Preprint No.206 (Spec. Astrophys. Observ., 2005).

\end{itemize}

\end{document}